\documentclass[twocolumn,showpacs,preprintnumbers,amsmath,amssymb]{revtex4}
\usepackage{graphicx}
\usepackage{dcolumn}
\usepackage{bm}
\begin{document}
\title {Bounds for the second and the third derivatives of the 
electron density at the nucleus}
\author {Zhixin Qian}%
\email{ZhixinQ@pku.edu.cn}
\affiliation{Department of Physics,
Peking University, Beijing 100871, China}
\date{\today}
\begin{abstract}
Lower bound for ${\bar \rho}''(0)$, the second derivative
of the spherically
averaged atomic electronic density at the nucleus, and upper bound for
${\bar \rho}'''(0)$,
the third derivative,
are obtained respectively. It is shown that, for
the ground state, 
${\bar \rho}''(0) \ge \frac{10}{3} Z^2 \rho(0)$ and
${\bar \rho}'''(0) \le -\frac{14}{3}Z^3 \rho(0)$ where
$Z$ is the charge of the nucleus, and $\rho(0)$ is 
the electron density
at the nucleus. 
Tighter bounds for ${\bar \rho}''(0)$ and
${\bar \rho}'''(0)$ are also reported which are
valid for both the ground state and excited states. 
Explicit illustration
with the example of one-electron atomic ions is given.
%These results
%hold for both ground-state and excited-state densities
%of many-electron systems. 
\end{abstract}
%\pacs{31.15.Ew, 31.10.+z, 71.10.-w}
\maketitle

\section{Introduction}

Rigorous knowledge of the electron density is rather valuable
in understanding the electronic structure of atoms, molecules,
and solids. It is also helpful in guiding the construction
of the approximations for the exchange-correlation 
energy functional and the corresponding potential in the
approach of density functional theory \cite{HKS} to 
the problems of inhomogeneous electron systems.
One of the well
known exact results is the so-called
Kato theorem \cite{kato} which states that, near any nucleus,
\begin{eqnarray}  \label{cusp}
{\bar \rho}'(r)|_{r=0} = -2Z \rho(0) .
\end{eqnarray}
Here $Z$ is the charge of the nucleus which is taken as the origin;
${\bar f}(r)$ means the spherical average of
function $f({\bf r})$;
primes denote the derivatives with respect to $r$
in this paper.

In recent work \cite{qian}, by
investigating the behavior of the wavefunctions of
the interacting Schr\"odinger system and the corresponding
noninteracting Kohn-Sham system in the vicinity of the nucleus, 
we established 
relations between the second 
and the third derivatives of the spherically averaged density
at the nucleus. They could be understood as extentions of the cusp
condition of Eq. (\ref{cusp}) to higher orders of derivatives.
%Several exact results for
%the Kohn-Sham exchange-correlation
%potential near the nucleus
%have been obtained from these relations in Ref. \cite{qian}. 
In
this paper, we derive rigorously a lower bound for the second
derivative and an upper bound 
for the third derivative, respectively,
of the spherically averaged density
at the nucleus for the ground state. The bounds are
given as follows:
\begin{eqnarray} \label{bound1}
{\bar \rho}''(0) \ge \frac{10}{3} Z^2 \rho(0),
\end{eqnarray}
and
\begin{eqnarray}  \label{bound2}
{\bar \rho}'''(0) \le -\frac{14}{3}Z^3 \rho(0).
\end{eqnarray}
%We note that these bounds hold for both ground-state
%and excited-state densities.
The results hold in general whether
the system is an atom, molecule or a solid.
Tighter bounds for ${\bar \rho}''(0)$ and
${\bar \rho}'''(0)$ are also reported 
(see Eqs. (\ref{inequ2}) and (\ref{inequ3})
in Sec. III) which
are valid for the excited states as well. 

In Sec. II, we discuss the near nucleus behavior of 
the wavefunction of the Schr\"odinger equation
and the density. The derivation for 
the bounds is presented in Sec. III.
%(\ref{bound1}) and (\ref{bound2}) is presented.
In Sec. IV explicit illustration
for the case of one-electron atomic ions is given.
Summarizing remarks are made in Sec. V.

\section{Near nucleus behavior of the wavefunction and the density}

The Schr\"odinger equation for $N$-electrons
in an external potential $v({\bf r})$ arising from
their interaction with nuclei is (in a.u.)
\begin{eqnarray} \label{schrodinger1}
{\hat H} \Psi = ({\hat T} + {\hat V} + {\hat U})\Psi = E \Psi ,
\end{eqnarray}
where ${\hat T}= - \frac{1}{2} \sum_{i} \bigtriangledown_i^2$,
${\hat V}= \sum_{i} v({\bf r}_i)$, 
${\hat U}= \frac{1}{2} \sum_{i \neq j}
\frac{1}{|{\bf r}_i - {\bf r}_j|}$, $\Psi$ the wavefunction, and $E$
the energy.
%The electron density is 
%defined as
%\begin{eqnarray} \label{density}
%\rho({\bf r}) = N \sum_s \int d{\bf x}_2 \dots d{\bf x}_N
%|\Psi({\bf r}s, {\bf x}_2, \dots {\bf x}_N)|^2,
%\end{eqnarray}
%where ${\bf x}_j$ denotes coordinate ${\bf r}_j$
%and spin $s_j$, and $\int d{\bf x}_j = \sum_{s_j} \int d{\bf r}_j$.
Following Ref. \cite{qian}, we write, for
limiting small $r$, the many-body wavefunction as
\begin{eqnarray}\label{wavefunction}
\Psi({\bf r}, {\bf X})&& =  \Psi(0, {\bf X}) + a({\bf X})r
+ b({\bf X})r^2 + c({\bf X}) r^3 + \dots   \nonumber \\
&& + \sum_{m=-1}^1[a_{1m}({\bf X})r
+ b_{1m}({\bf X})r^2 +\dots] Y_{1m}({\hat r})  \nonumber \\
&& + \sum_{m=-2}^2 [b_{2m}({\bf X})r^2 
+ \dots] Y_{2m}({\hat r}) \nonumber \\
&&+ \dots,
\end{eqnarray}
where ${\hat r}={\bf r}/r$, ${\bf X}$
denotes $s, {\bf r}_2 s_2, \dots, {\bf r}_N s_N$,
and
$Y_{lm}({\hat r})$ are the spherical harmonics. For $r \to 0$,
the Schr\"odinger equation (\ref{schrodinger1})
can be rewritten as \cite{qian}
\begin{eqnarray} \label{schrodinger2}
[&-&\frac{1}{2} \bigtriangledown^2 -\frac{Z}{r}
+ r \sum_{m=-1}^{1}Y_{1m}({\hat r})
g_m({\bf X})] \Psi({\bf r}, {\bf X})  \nonumber \\
&+&H_{Z-1}^{N-1}({\bf X}) \Psi({\bf r}, {\bf X})
= E \Psi({\bf r}, {\bf X}) ,
\end{eqnarray}
where
\begin{eqnarray}
g_m({\bf X}) = \frac{4 \pi}{3} \sum_{i=2}^N \frac{1}{r_i^2}
Y_{1m}^*({\hat r}_i),
\end{eqnarray}
and
\begin{eqnarray} 
H_{Z-1}^{N-1}({\bf X})=&& \sum_{i=2}^N 
[-\frac{1}{2}\bigtriangledown_i^2
+ v({\bf r}_i) + \frac{1}{r_i} ]  \nonumber \\
&& + \frac{1}{2} \sum_{i \neq j \neq 1}^N
\frac{1}{|{\bf r}_i - {\bf r}_j|}.  
\end{eqnarray}
Substituting the expression of Eq. (\ref{wavefunction}) 
into Eq. (\ref{schrodinger2}) and equating 
the coefficients of the terms of $r^{-1}$,
$r^0 Y_{1m}({\hat r})$, $r^0$, and $r^1$, respectively, one has
\begin{eqnarray}\label{a-Psi(0)}
&&a({\bf X}) + Z \Psi(0, {\bf X})=0 ,  \\
&&2 b_{1m}({\bf X}) + Za_{1m}({\bf X}) =0 ,  \\
&&3 b({\bf X}) - Z^2 \Psi(0, {\bf X}) = [H_{Z-1}^{N-1}({\bf X})-E]
\Psi(0, {\bf X}), \\ 
&&6 c({\bf X}) + Z b({\bf X}) = Z[E-H_{Z-1}^{N-1}({\bf X})]
\Psi(0, {\bf X}). 
\end{eqnarray}
We note that the behavior of the wavefunction in the vicinity
of the nucleus has
been extensively investigated \cite{kato,qian,cusp1}.
In fact, Eqs. (9) and (10) have been obtained previously.
On the other side, Eqs. (11) and (12), which shall play a key role
in the derivation in Sec. III, have not been previously reported
{\em in this separate form}.
% Similar
%relations have been obtained previously, but not in the
%above form of Eqs. (9-12). Particularly, Eqs. (11) and (12)
%which are absent in the literature shall play a key role
%in the derivation in Sec. III. 

The density  
is defined as
%respectively, as
\begin{eqnarray}   \label{density}
\rho({\bf r})= N \int d{\bf X} |\Psi({\bf r}, {\bf X})|^2 ,
\end{eqnarray}
%and
%\begin{eqnarray}  \label{t-definition}
%t({\bf r}) =
%\frac{1}{2} N \int d {\bf X} \bigtriangledown \Psi^*({\bf r}, {\bf X})
%\cdot \bigtriangledown \Psi({\bf r}, {\bf X}),
%\end{eqnarray}
where $\int d{\bf X}$ denotes $\sum_s
\int d{\bf x}_2 \dots d{\bf x}_N$.
With the relations in Eqs. (9) and (10), it can be shown that 
\begin{eqnarray}   \label{second}
{\bar \rho}''(0) = && 2Z^2 \rho(0) + 
4N \int d{\bf X}
Re[\Psi^*(0, {\bf X}) b({\bf X})]  \nonumber \\ 
%&& + \frac{2}{3} [2{\bar t}(0) - Z^2 \rho(0)],
&& + 2N \int d {\bf X} \sum_{m=-1}^{1}
\frac{1}{4 \pi} |a_{1m} ({\bf X})|^2
\end{eqnarray}
and
\begin{eqnarray}   \label{third}
{\bar \rho}'''(0) =&& 12N \int d{\bf X}
Re[\Psi^*(0, {\bf X})(c({\bf X})-Zb({\bf X}))] \nonumber \\
&& - 6ZN \int d {\bf X} \sum_{m=-1}^{1}
\frac{1}{4 \pi} |a_{1m} ({\bf X})|^2,
\end{eqnarray}
(see also Ref. \cite{qian}.)
This completes the discussion of the behavior of
the wavefunction and the density near the nucleus.
% for the derivation
%of Eqs. (\ref{bound1}) and
%(\ref{bound2}).

\section{Derivation of Eqs. (\ref{bound1}) and 
(\ref{bound2})}

%To establish the bounds for the derivatives of the
%spherically averaged density, we 
We observe that Eqs. (11), (12) and Eq. (\ref{density}) lead to
\begin{eqnarray} \label{keystep1}
N  \int d {\bf X} && Re
[\Psi^*(0, {\bf X}) b({\bf X})]  \nonumber \\
\ge && \frac{1}{3} ( Z^2 -E     
+ E_{Z-1,0}^{N-1}) \rho(0), 
\end{eqnarray}
\begin{eqnarray} \label{keystep2}
N  \int d {\bf X} && Re
[\Psi^*(0, {\bf X}) c({\bf X})]  \nonumber \\
\le &&  \frac{1}{18}Z(-Z^2 +4E - 4E_{Z-1,0}^{N-1}) \rho(0),
\end{eqnarray}
where $E_{Z-1,0}^{N-1}$ is the ground state energy of 
the Hamiltonian $H_{Z-1}^{N-1}({\bf X})$. 
Inequalities (\ref{keystep1}) and (\ref{keystep2}) are 
critical equations in
the following derivation.
Comparing them with Eq. (\ref{second}) 
and Eq. (\ref{third}), respectively, 
one has
\begin{eqnarray} \label{second2}
{\bar \rho}''(0) \ge && \frac{10}{3}Z^2 \rho(0) +
\frac{4}{3} (E_{Z-1,0}^{N-1}
-E) \rho(0)    \nonumber \\
%&& +\frac{2}{3} [2{\bar t}(0) -Z^2 \rho(0)],
&& + 2N \int d {\bf X} \sum_{m=-1}^{1}
\frac{1}{4 \pi} |a_{1m} ({\bf X})|^2,
\end{eqnarray}
and
\begin{eqnarray}   \label{third2}
{\bar \rho}'''(0) \le && -\frac{14}{3}Z^3 \rho(0)
- \frac{20}{3}Z(E_{Z-1,0}^{N-1}
-E) \rho(0)  \nonumber \\
%&& -2Z [2{\bar t}(0) -Z^2 \rho(0)].
&&  - 6ZN \int d {\bf X} \sum_{m=-1}^{1}
\frac{1}{4 \pi} |a_{1m} ({\bf X})|^2.
\end{eqnarray}

In passing, we mention that, at the nucleus, 
the kinetic energy density,
which is defined as
\begin{eqnarray}  \label{t-definition}
t({\bf r}) =
\frac{1}{2} N \int d {\bf X} \bigtriangledown \Psi^*({\bf r}, {\bf X})
\cdot \bigtriangledown \Psi({\bf r}, {\bf X}),
\end{eqnarray}
has been shown in Ref. \cite{qian} as
\begin{eqnarray} \label{t(0)}
{\bar t}(0) -\frac{1}{2} Z^2 \rho(0) 
= N \int d {\bf X} \sum_{m=-1}^{1}
\frac{3}{8 \pi} |a_{1m} ({\bf X})|^2 .
\end{eqnarray}
Equation (\ref{t(0)}) indicates that
\begin{eqnarray}  \label{t(0)2}
{\bar t}(0) \ge \frac{1}{2} Z^2 \rho(0).
\end{eqnarray}
%It might be worth mentioning again \cite{qian} that the
The existence of the term on the right hand side of Eq. (\ref{t(0)})
had not been recognized before \cite{dreizler}.
%By using Eq. (\ref{t(0)2}) in 

From Eqs. (\ref{second2}) and (\ref{third2}),
one obtains
\begin{eqnarray}   \label{inequ2}
{\bar \rho}''(0) \ge \frac{10}{3}Z^2 \rho(0) 
+ \frac{4}{3}(E_{Z-1,0}^{N-1}
-E) \rho(0),
%\nonumber \\
%&&\ge \frac{10}{3}Z^2 \rho(0)
\end{eqnarray}
and
\begin{eqnarray}  \label{inequ3}
{\bar \rho}'''(0)  \le -\frac{14}{3}Z^3 \rho(0)
- \frac{20}{3}Z(E_{Z-1,0}^{N-1}
-E) \rho(0).
%\nonumber \\
%&&\le -\frac{14}{3}Z^3 \rho(0)
\end{eqnarray}
Up to this point, all the calculations 
are in fact not restricted to the ground
state. The bounds
shown in Eqs. (\ref{inequ2}) and (\ref{inequ3}) are valid for
{\em both the ground state and excited states}. For the ground
state
% which we are concerned with in this paper, 
%by the use of the fact that $E_{Z-1,0}^{N-1}-E \ge 0$, we
we further obtain Eqs. (\ref{bound1}) and (\ref{bound2})
from Eqs. (\ref{inequ2}) and (\ref{inequ3}).

\section{Illustration for one-electron atomic ions}

For one-electron atomic ions (including
hydrogen atom),
%the wavefunctions are exactly known for both the ground
%state and the excited states. In this case,
Eqs. (\ref{inequ2}) and (\ref{inequ3}) become
\begin{eqnarray}   \label{inequ22}
{\bar \rho}''(0) \ge \frac{2}{3}(5Z^2
-2E) \rho(0),
\end{eqnarray}
and
\begin{eqnarray}  \label{inequ33}
{\bar \rho}'''(0)  \le -\frac{2}{3}Z
(7Z^2-10E) \rho(0).
\end{eqnarray}
The wavefunctions are exactly known for both the ground
state and excited states:
$\Psi({\bf r}) =
R_{nl}(r) Y_{lm}({\hat r})$, with  \cite{schiff}
\begin{eqnarray} \label{one-wf}
R_{nl}(r)= - && \biggl \{ \alpha_n^{3+2l}
\frac{(n-l-1)!}{2n [(n+l)!]^3} \biggr \}^{1/2}
e^{-\alpha_nr/2}    \nonumber \\
&& r^{l}  L_{n+l}^{2l+1}(\alpha_nr),
\end{eqnarray}
where $\alpha_n=2Z/n$, and $L_{n+l}^{2l+1}(\alpha_nr)$
are the Laguerre polynomials.
It is easy to see that for $l \neq 1$, $a_{1m}=0$
in Eqs. (\ref{second2}) and (\ref{third2}), and
Eqs. (\ref{inequ2}) and (\ref{inequ3}) in fact become 
equalities:
\begin{eqnarray}   \label{equ22}
{\bar \rho}''(0) = \frac{2}{3}(5Z^2
-2E) \rho(0),
%\nonumber \\
%&&\ge \frac{10}{3}Z^2 \rho(0)
\end{eqnarray}
and
\begin{eqnarray}  \label{equ33}
{\bar \rho}'''(0)  = -\frac{2}{3}Z
(7Z^2-10E) \rho(0).
%\nonumber \\
%&&\le -\frac{14}{3}Z^3 \rho(0)
\end{eqnarray}

We next give an explicit illustration of the results
with the exactly known wavefunctions.
Obviously for $l \ge 2$ Eqs. (\ref{equ22}) and (\ref{equ33})
are trivially true since
all $\rho(0)$, ${\bar \rho}''(0)$, and ${\bar \rho}'''(0)$
are zero. For $l=0$, one can obtain from Eq. (\ref{one-wf})
\begin{eqnarray}
{\bar \rho}(r)=\frac{Z^3}{\pi n^3}
\biggl [ 1&-& 2Zr + \frac{1}{3}Z^2(5
+\frac{1}{n^2}) r^2  \nonumber \\
&-&\frac{1}{9}Z^3(7+\frac{5}{n^2}) r^3 + \dots 
\biggr ] .
\end{eqnarray}
Equations (\ref{equ22})
and (\ref{equ33}) are hence confirmed with the help
of the fact that $E=-Z^2/2n^2$. 
For $l=1$, one has
\begin{eqnarray}
{\bar \rho}(r)=\frac{1}{9 \pi} \biggl (
\frac{Z}{n} \biggr )^5 (n^2 -1) r^2 (1 - Zr 
+\dots ).
\end{eqnarray}
The inequalities of Eqs. (\ref{inequ22}) and (\ref{inequ33})
are satisfied. Notice that in this case $a_{1m} \neq 0$ 
in Eqs. (\ref{second2}) and (\ref{third2}).

\section{summary}

In summary, in this paper, we have established
the lower bound for ${\bar \rho}''(0)$, the second derivative of 
the spherically averaged 
electron density at the nucleus, 
in Eq. (2), and
the upper bound for ${\bar \rho}'''(0)$, the third derivative, 
in Eq. (3). Tighter bounds for ${\bar \rho}''(0)$
and ${\bar \rho}'''(0)$, valid for
both the ground state and excited states, are also reported in
Eqs. (\ref{inequ2}) and (\ref{inequ3}).
These results add some to our rigorous information of
the electron density, which might be valuable in the 
calculation of the electronic structure of atomic systems.
%such as those based on
%density functional theory. 
%We finally mention that
%these inequalities hold also for excited states, for
%the derivation is evidently not restricted only to
%the ground state. 
%They are valid in general whether
%the system is an atom, molecule or a solid.

Whether the opposite bounds exist for ${\bar \rho}''(0)$
and ${\bar \rho}'''(0)$  remains an interesting question.
Not unrelated to this question is the fact 
that both upper and lower bounds have
been extensively explored for $\rho (0)$ 
(and, according to Eq. (\ref{cusp}), 
equally for ${\bar \rho}'(0)$) \cite{hoffmann}.
%according to Eq. (\ref{cusp}) \cite{hoffmann}

{\em Note added.} During the preparation of this manuscript, 
the author became aware of 
related work \cite{fournais}.  
The bounds shown in Eqs. (\ref{bound1})
and (\ref{bound2}) in this paper are indeed the same
as those in
Eqs. (1.25) and (1.32) in Ref. \cite{fournais}
(up to a different form of Hamiltonians).
%are in agreement with Eqs. (\ref{bound1})
%and (\ref{bound2}) of this paper.
% and the coefficient
%differences arise from the different definitions
%of the Hamiltonian, i.e., the absence of the factor of $1/2$ 
%in front of
%the kinetic energy operator in Eq. (1.1) in \cite{fournais}. 
There are, however, minor 
differences
%, however, exist in the bounds 
%established in the two papers.
between inequalities (\ref{inequ2}) and (\ref{inequ3})
of the present paper and the first inequalities
of Eqs. (1.25)
and (1.32) in Ref. \cite{fournais}.
In fact,  $E_{Z-1,0}^{N-1}
-E$ can be rewritten as $E_{Z-1,0}^{N-1} - E_{Z,0}^{N-1} - \mu$,
where $\mu= E -E_{Z,0}^{N-1}$ is the minus
the ionization energy and denoted as $-\epsilon$
in Eq. (1.20) in Ref. \cite{fournais}. Since 
$E_{Z-1,0}^{N-1} - E_{Z,0}^{N-1} \ge 0$, it is easy to see that 
the bounds in Eqs. (\ref{inequ2}) and (\ref{inequ3})
are tighter than those given by the first inequalities of Eqs. (1.25) 
and (1.32). The differences were also recognized and analyzed
in Remark 1.4 in Ref. \cite{fournais}, though equations like
inequalities (\ref{inequ2}) and (\ref{inequ3}) were not
explicitly given there.

% in \cite{fournais}.

The author is indebted to
Profs. S. Fournais,
M. Hoffmann-Ostenhof, T. Hoffmann-Ostenhof,
and T. {\O}stergaard S{\o}rensen
for confirming that the bounds shown in Eqs. (\ref{bound1})
and (\ref{bound2}) are the same
as those in
Eqs. (1.25) and (1.32) in Ref. \cite{fournais}. 
Enlightening 
comments on this paper from these colleagues 
are also gratefully acknowledged.
This work was supported by the National Science Foundation
of China under Grant No. 10474001.

\end{document}